\documentclass{jnmp_fc}
\usepackage[dvips]{graphicx}
\usepackage{amsmath}
\JNMPnumberwithin{equation}{section}
\begin{document}
\renewcommand{\evenhead}{C Jung, T H Seligman and J M Torres}
\renewcommand{\oddhead}{Canonically Transformed Detectors Applied}
\thispagestyle{empty}
\FirstPageHead{12}{1}{2005}
{\pageref{firstpage}--\pageref{lastpage}}{{\bf \tiny
{Birthday Issue}}}

\copyrightnote{2005}{C Jung, T H Seligman and J M Torres}
\Name{Canonically Transformed Detectors Applied to the Classical Inverse
Scattering Problem}
\label{firstpage}
\Author{C JUNG, T H SELIGMAN and J M TORRES}
\Address{Centro de  Ciencias   F\'isicas,  
Universidad Nacional  Aut\'onoma de M\'exico, Cuernavaca, Morelos, M\'exico\\
~~E-mail: jung@fis.unam.mx, seligman@fis.unam.mx, mau@fisica.unam.mx}

\Date{This article is part of the special issue published
in honour of Francesco Calogero on the occasion of his 70th birthday}

\begin{abstract}
\noindent
The concept  of measurement in classical scattering  is interpreted as
an overlap  of a particle  packet with some  area in phase  space that
describes the detector. Considering that usually we record the passage
of particles  at some point in  space, a common  detector is described
{\it  e.g.} for one-dimensional systems  as a  narrow  strip in  phase
space.   We  generalize  this   concept  allowing  this  strip  to  be
transformed  by some,  possibly non-linear,  canonical transformation,
introducing  thus a  canonically  transformed detector.  We show  such
detectors  to be  useful  in  the context  of  the inverse  scattering
problem  in  situations where  recently  discovered scattering  echoes
could not be  seen without their help.  More  relevant applications in
quantum systems are suggested.
\end{abstract}

\section{Introduction}


 Usually  a classical  scattering measurement  will determine  at what
time  the outgoing particle  passes a  certain point  in a  surface of
codimension  one in  configuration space,  which in  one  dimension is
simply characterized by  a distance from the center  of the scattering
region, and in more dimensions will also include scattering angles. If
we allow for a non-zero thickness of the detector this will describe a
region  in  configuration  space.  If  in addition  we  consider  that
normally  particles can  be detected  only within  some  finite energy
range, this in turn implies a finite region in phase space.  Depending
on  the  characteristics  of  a  real  experiment  or  on  theoretical
advantages  for a  one-dimensional system  it could  be  any structure
narrow in configuration space and  rather wider in momentum space.  We
now consider an experiment where  the incoming particles form a narrow
packet in phase space. The result  of this experiment will be given as
the  time evolution  of the  overlap in  phase space  of  the evolving
packet with the phase space strip characterizing the detector.

The purpose  of the present paper  is to generalize this  concept of a
detector by  allowing canonical transformations  of the area  in phase
space that describes the detector, in  order to be able to detect more
subtle structures in  the outgoing particle packet. We  shall see that
such  generalized  detectors  are  essential to  solve  the  classical
inverse scattering problem in  certain cases, and we shall demonstrate
the usefulness  of the  concept in a  simple example, which  will also
provide typical transformations needed.

In two recent  publications \cite{e1,e2} it has been  suggested to use
the self-pulsing effect of chaotic scattering ( scattering echoes ) to
solve the  chaotic inverse scattering problem.  Recently an experiment
in  superconductive   cavities  \cite{darmstadt}  has   confirmed  the
theoretical prediction \cite{e1,e2}. The basic idea is the following:

Imagine a  scattering system with  two essential degrees of  freedom (
one of  them may be a  periodic explicit time dependence,  the case we
are  mainly  concerned  with in  the  present  paper  ) which  can  be
converted into  a 2-dimensional iterated map, either  a Poincar\'e map
or a stroboscopic map. We consider the case where this map has a large
KAM island  surrounded by a  homoclinic/heteroclinic tangle.  Incoming
particles form a narrow packet in phase space. Many of these particles
undergo a  direct scattering process and emerge  immediately. The rest
comes in  contact with the homoclinic/heteroclinic  tangle and rotates
around the  KAM island.  After  each complete rotation some  leave the
homoclinic tangle and produce an  outgoing particle pulse. We call the
sequence of outgoing pulses echoes. The time difference $\tau$ between
adjacent echoes  is determined by the  time it takes  to circle around
the KAM island and this time in turn is given by the development stage
of  the homoclinic tangle.  For example  for the  case of  a symmetric
ternary horseshoe, which we shall encounter later, this development is
described  by a  parameter  $\gamma$.  It varies  between  zero for  a
separatrix line and one for a complete horseshoe \cite{i1} and relates
to the rotation period $T$ as
\begin{equation}\label{develop}
 T = - 2 log_3(\gamma) + 3/2
\end {equation}
Thereby  in the  end a  measurement of  the temporal  distance between
adjacent echoes gives the development degree of the homoclinic tangle.

The  crucial  point is  a  clean measurement  of  the  echoes. If  all
particles in  any echo  would have exactly  the same speed,  then each
packet would stay  together and we could measure  them at any distance
from the target. However, in reality  each echo consists of a group of
particles  with  different  speeds  and  therefore  the  packets  have
dispersion {\it i.e.} become broader with increasing distance from the
target. At some  distance the echoes become broader  than the distance
between two adjacent ones. A separation into a sequence of consecutive
echoes is  no longer  possible. At  first one could  think of  using a
detector which only  accepts a small velocity window.  This might work
if the velocity spread is  not too large. However it becomes difficult
because of  two reasons. First,  for attractive potentials,  where the
outer fixed point  of the Poincar\'e map is  located at infinity, some
of the outgoing  particles have velocities close to  zero and then the
relative spread  of velocities  is beyond any  limit. Second,  a small
velocity  window of  the detector  only accepts  a small  part  of the
particles and thereby leads to a very small signal strength. There are
two completely different strategies  to circumvent these problems. One
possibility is  a measurement  of the energy  dependence of  the cross
section and  to Fourier transform from  the energy domain  to the time
domain (see discussion in \cite{e2, darmstadt}).

If we  insist to  measure the sequence  of echoes directly  the second
possibility is a detector  whose sensitivity depends on an appropriate
combination  of   position  and   velocity,  i.e.  a   detector  whose
sensitivity  covers an  appropriate strip  in  phase space  or in  the
domain of  the Poincar\'e  map. The  ideal case is  to cover  a single
unstable tendril  of the  homoclinic tangle. This  can be  achieved by
canonically  transforming the  phase  space strip  that describes  the
simple minded (inadequate) detector.  We shall call this a canonically
transformed detector (CTD). In  one approach to the inverse scattering
problem,  discussed in  \cite{i1,i2}, a  central point  is to  find to
which hierarchical  level a  particular scattering trajectory  (or the
corresponding   interval  of   continuity  of   scattering  functions)
belongs.  In the  present setup  the answer  to this  question becomes
almost  trivial. If  we place  the set  of initial  conditions  into a
single  stable tendril  and  the  detector is  sensitive  in a  single
unstable tendril only, then the total delay time (measured in units of
the period of the kick) is  the hierarchical level up to an irrelevant
global constant.

In  the present paper  we concentrate  for simplicity  on periodically
kicked systems in a  1-dimensional position space where the homoclinic
tangle extends  to infinity. In the  next section we  shall discuss an
example  giving rise  to a  symmetric  ternary horseshoe,  and in  the
following  section we  shall show  how a  CTD can  detect  echoes that
cannot be seen in the usual way. Some concluding remarks will complete
our argument.

\section{The system and its homoclinic tangle}

\begin{figure}[!t]
\begin{center}
\includegraphics[scale=.9]{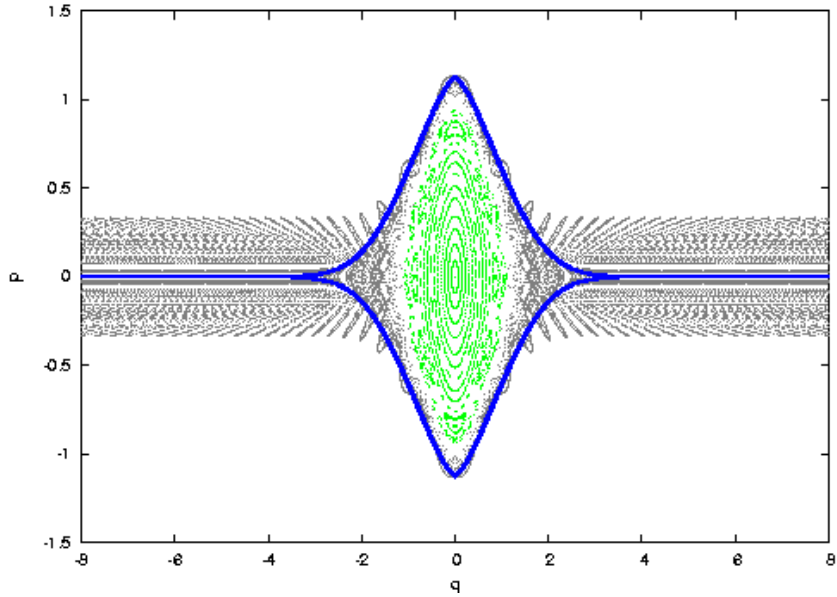}
\caption{\label{hor} Horseshoe of  low development in the stroboscopic
map given by Eq.~\ref{map}.  KAM  tori and secondary islands in green,
the boundary of  the fundamental rectangle R in  blue and the tendrils
of the invariant manifolds of the outer fixed points in grey.}
\end{center}
\end{figure}

We use the system described by the Hamiltonian
\begin{equation}\label{ham}
H(q,p,t)=\frac{p^{2}}{2}+A\,\,
V(q)\sum_{n=-\infty}^\infty\delta(t-n).
\end{equation}
Here $q$ is the position and  $p$ the momentum of the particle at time
$t$. $V(q)=-exp(-q^2)$  indicates the  potential function of  the kick
and  $A$  its strenght.  It  is a  periodically  kicked  system on  an
one-dimensional   configuration  space   with   a  purely   attractive
potential. We represent the dynamics  by the stroboscopic map taken at
times  $t =  n +1/2$,  {\it  i.e.} always  in the  middle between  two
consecutive  kicks.  This  choice  is  adapted to  the  time  reversal
symmetry of  our system in  so far as  time reversal corresponds  to a
reflection in the  $q$ axis and this reflection  maps stable manifolds
into unstable  manifolds and  vice versa. The  map is given  in closed
form as
\begin{equation}\label{map}
\begin{array}{l}
p_{n+1}=p_{n}-AV'(q_{n}+p_{n}/2)\\
q_{n+1}=q_{n}+\frac{1}{2}(p_{n}+p_{n+1})
\end{array}
\end{equation}
Because the  potential is purely  attractive the system does  not have
any point  of no return  in configuration space.   Therefore recurrent
trajectories are not confined to  a finite part of configuration space
and  also the  chaotic invariant  set (homoclinic  tangle)  extends to
infinity. The outermost  fixed points of the horseshoe  are located at
infinity.  Since  in the asymptotic  region the dynamics  is invariant
under   translations   in  position   this   outer   fixed  point   is
parabolic. However, under inclusion  of nonlinearities of the dynamics
it has a stable and an unstable invariant manifold reaching inwards to
the interaction region. The local branch of the stable manifold of the
point  at infinity  consists  of trajectories  which  run in  position
monotonically to  infinity while the  momentum converges to  zero. The
corresponding  part  of the  unstable  manifold  is  obtained by  time
reversal.   The  horseshoe is  traced  by  the  continuation of  these
invariant manifolds of the points at infinity. Figure ~\ref{hor} shows
the stroboscopic map  Eq.~\ref{map}. In the center marked  in green we
see a  few KAM tori  around the central  fixed point at the  origin as
well  as  a  few  secondary  islands, all  inside  the  outermost  KAM
surface. In blue we mark the boundary of the fundamental rectangle $R$
\cite{i1,i3}  that extends  to infinity  for zero  momenta.   The grey
points are further iterations of  the blue line and they show tendrils
in the asymptotic  region as well as in the  chaotic layer between the
outermost  KAM  surface  and  the  boundary of  $R$.   This  layer  is
interspersed with a fractal set  of integrable islands, the largest of
which appear as white areas within this layer.
A small  strip around the line of  momentum zero in the  domain of the
map (its boundaries are the local segments of the invariant manifolds)
belongs to  the fundamental  region $R$ of  the horseshoe even  for $q
\rightarrow \pm \infty$. Therefore we  exclude a larger strip from the
asymptotic region and  define the asymptotic region in  phase space as
the subset  of the domain which,  first, is far out  in position space
such that  the potential  is close to  zero and, second,  the absolute
value of the momentum is large  compared to the momentum values of the
boundary of $R$.
\begin{figure}[!t]
\begin{center}
\includegraphics[scale=.4]{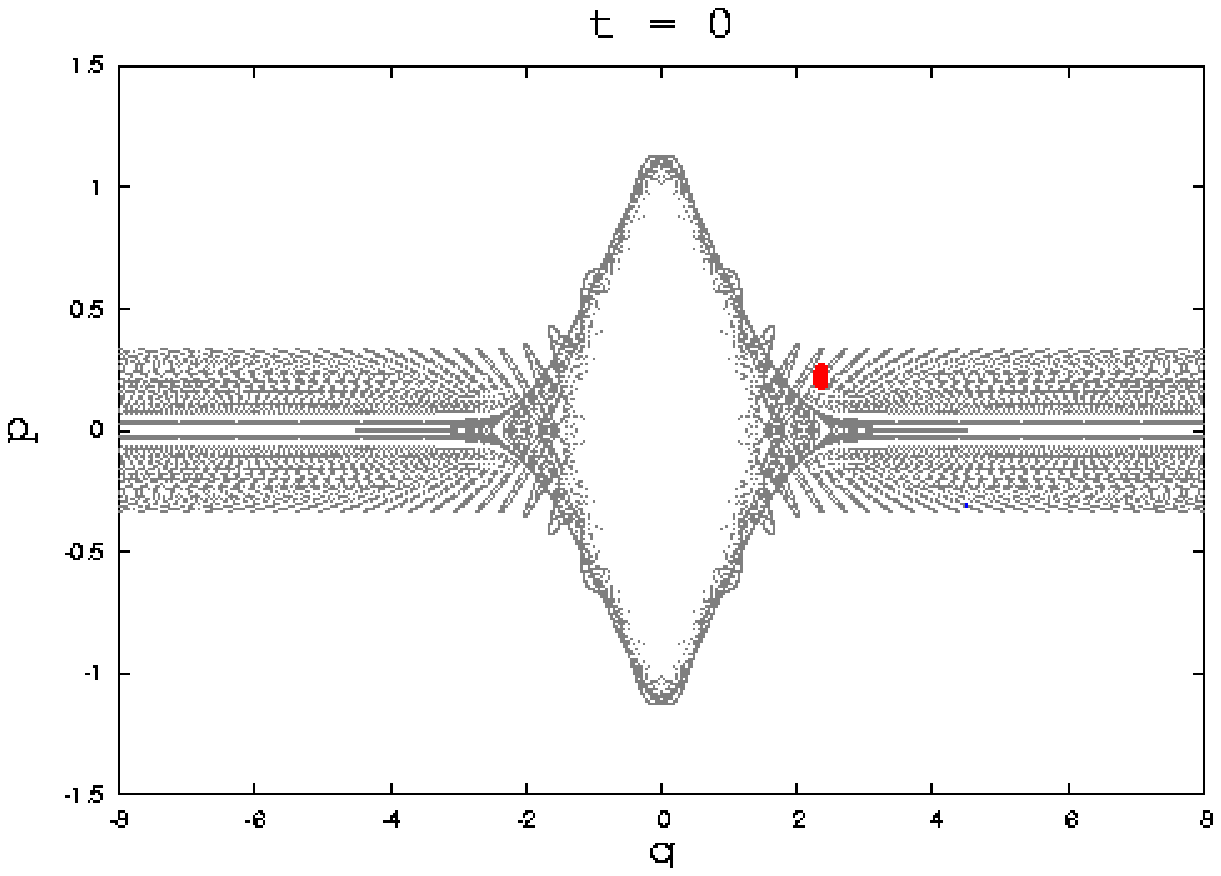}
\includegraphics[scale=.4]{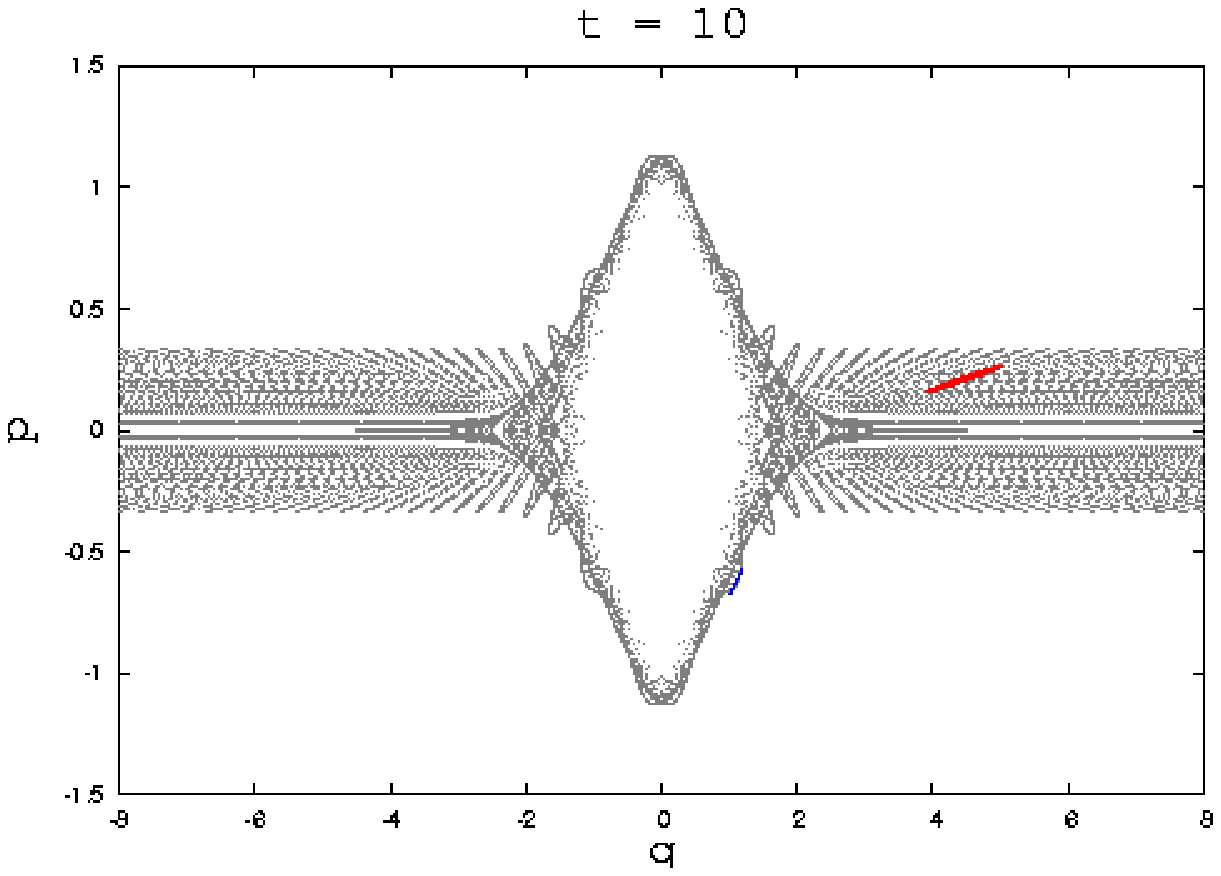}
\includegraphics[scale=.4]{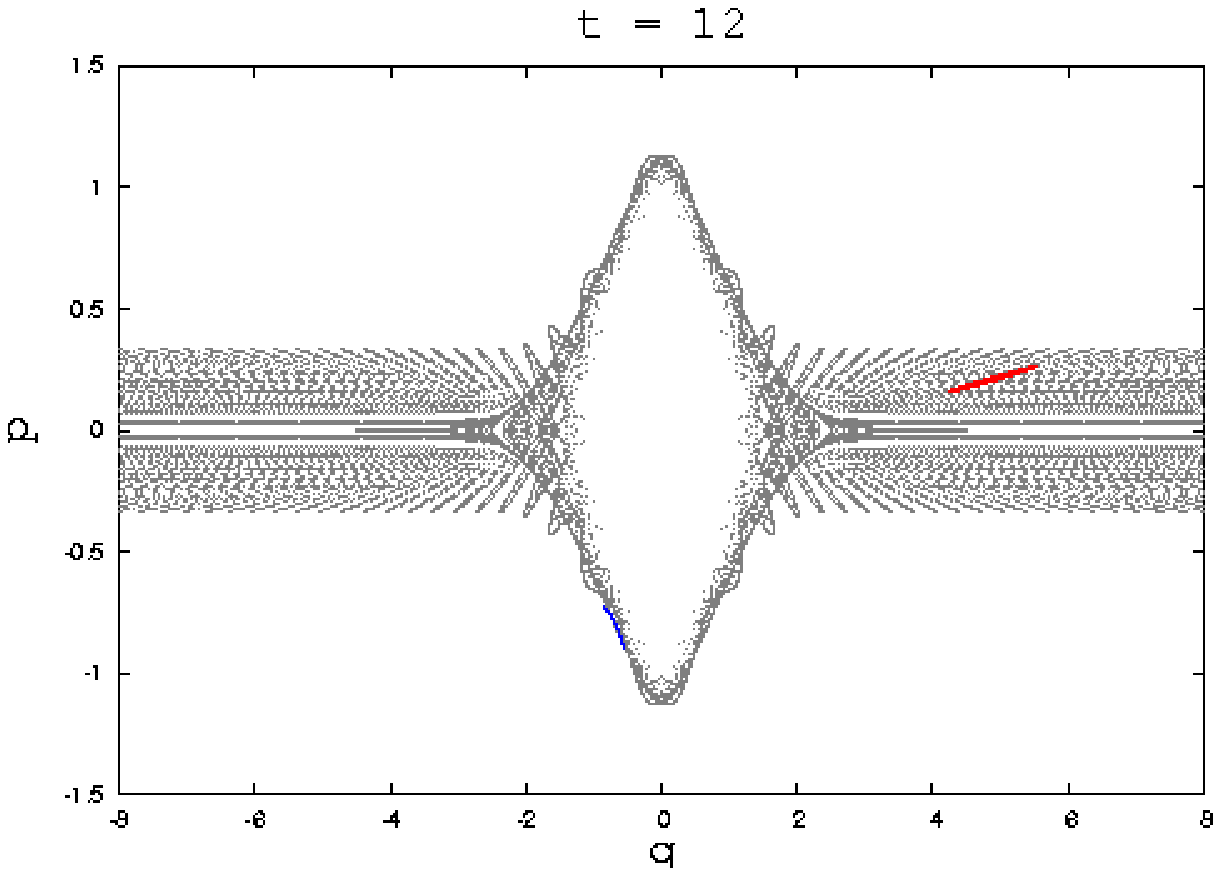}
\includegraphics[scale=.4]{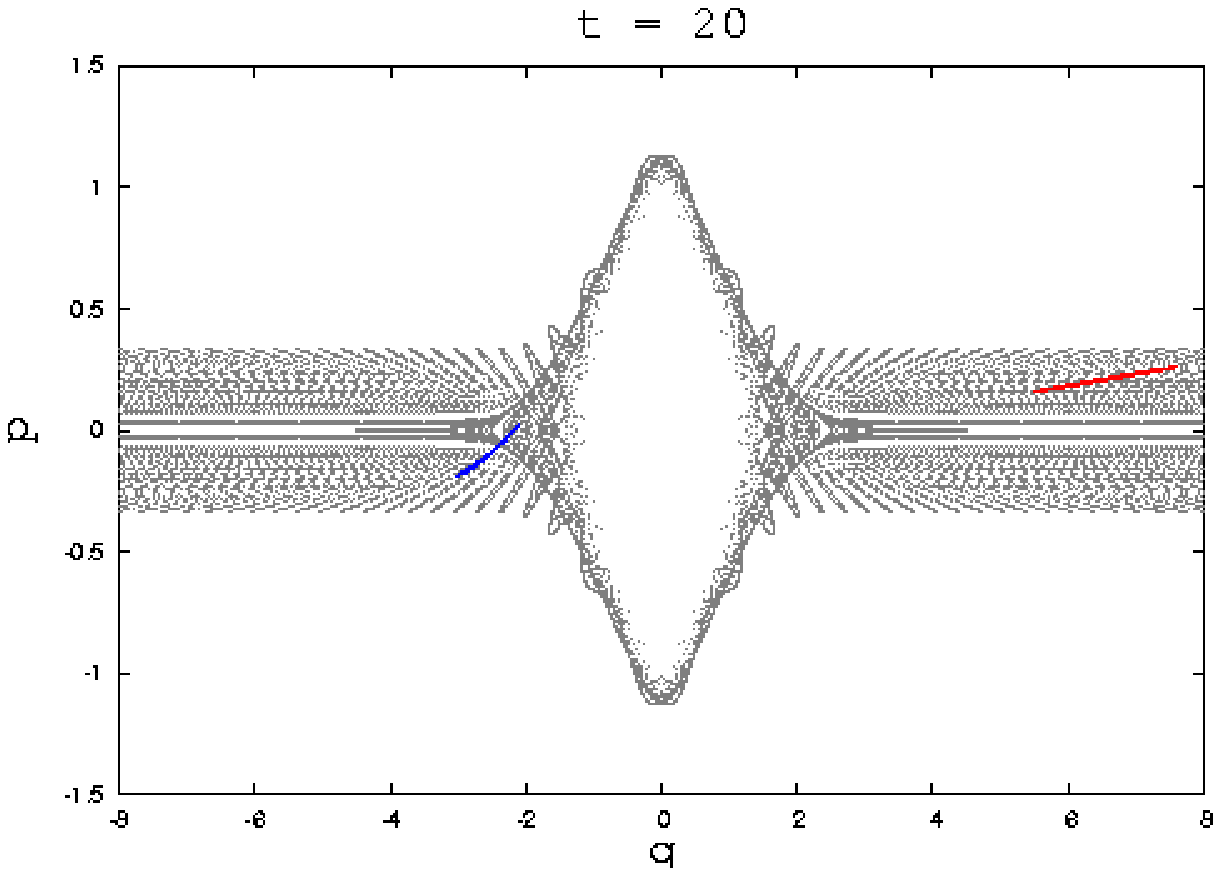}
\includegraphics[scale=.4]{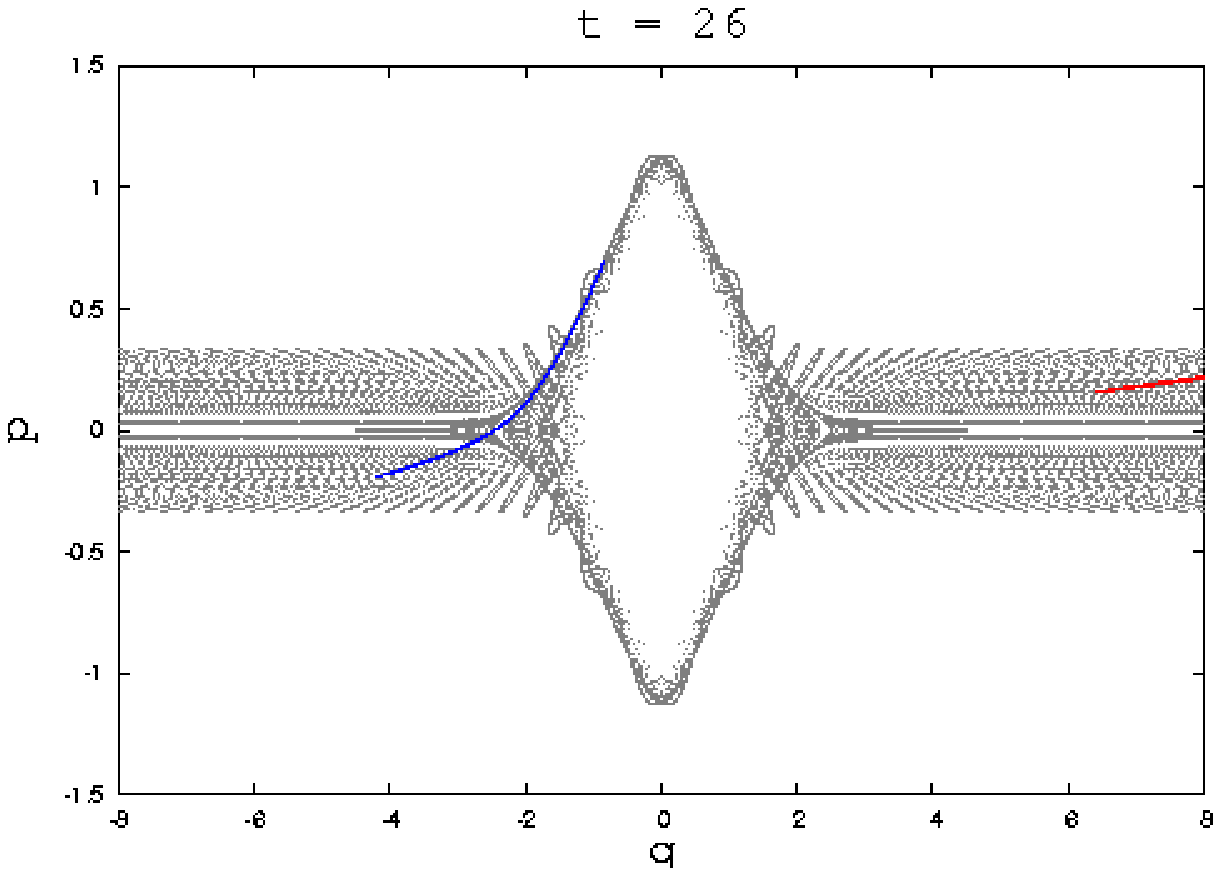}
\includegraphics[scale=.4]{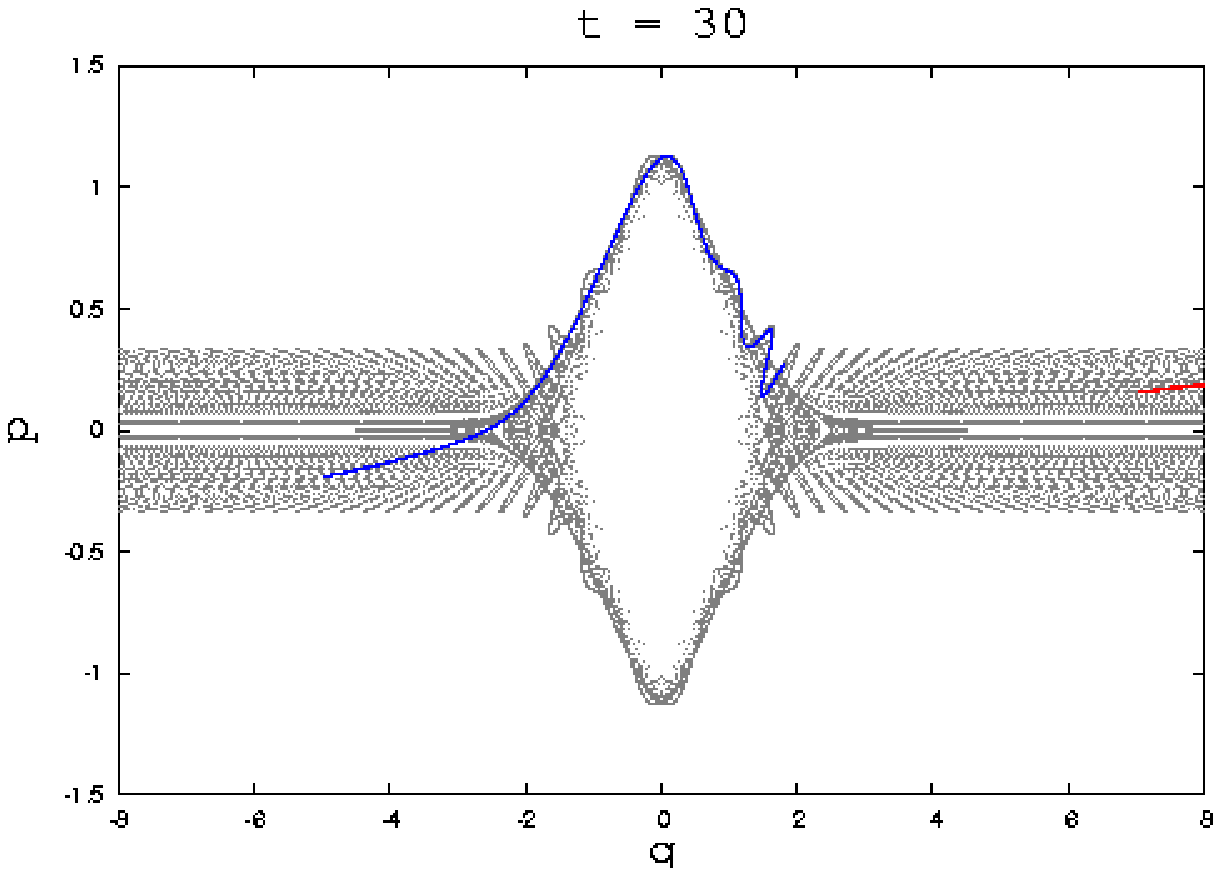}
\includegraphics[scale=.4]{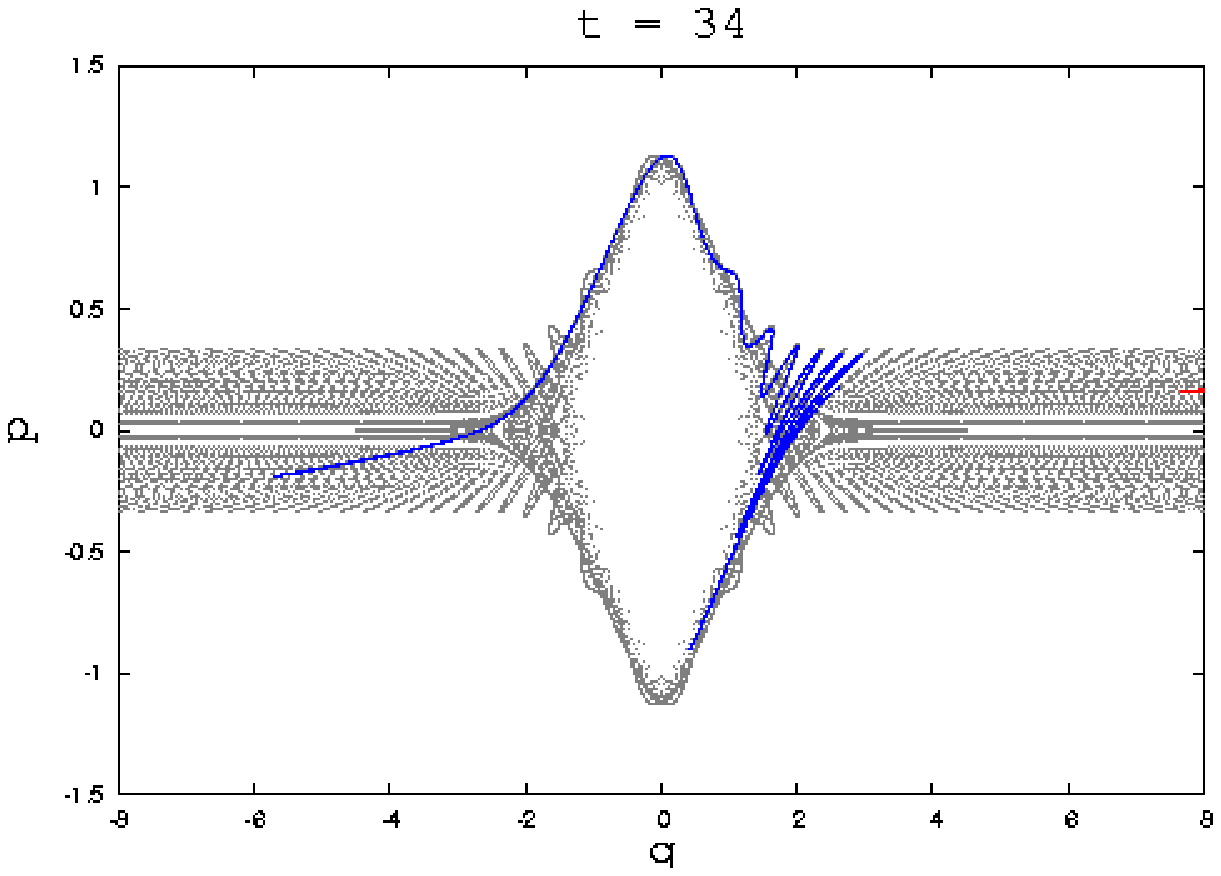}
\includegraphics[scale=.4]{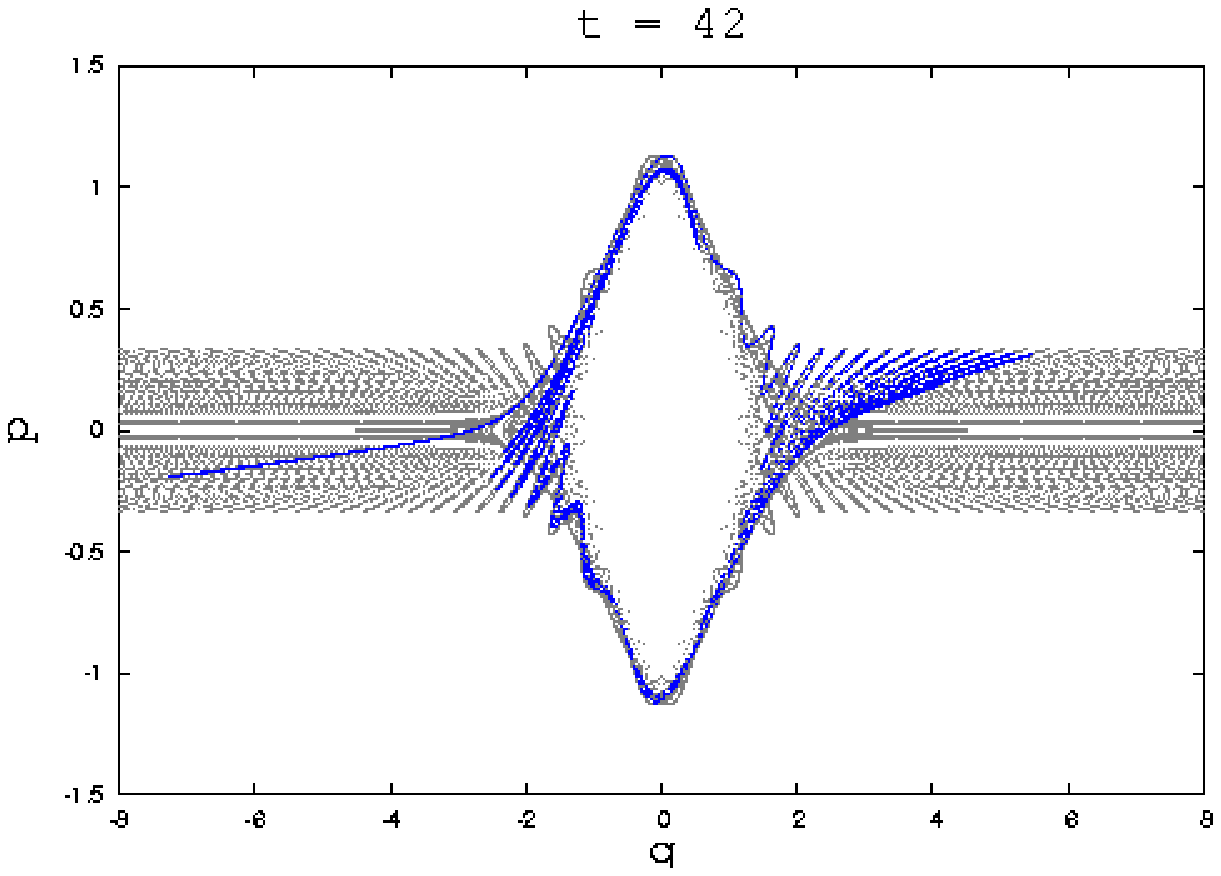}
\includegraphics[scale=.4]{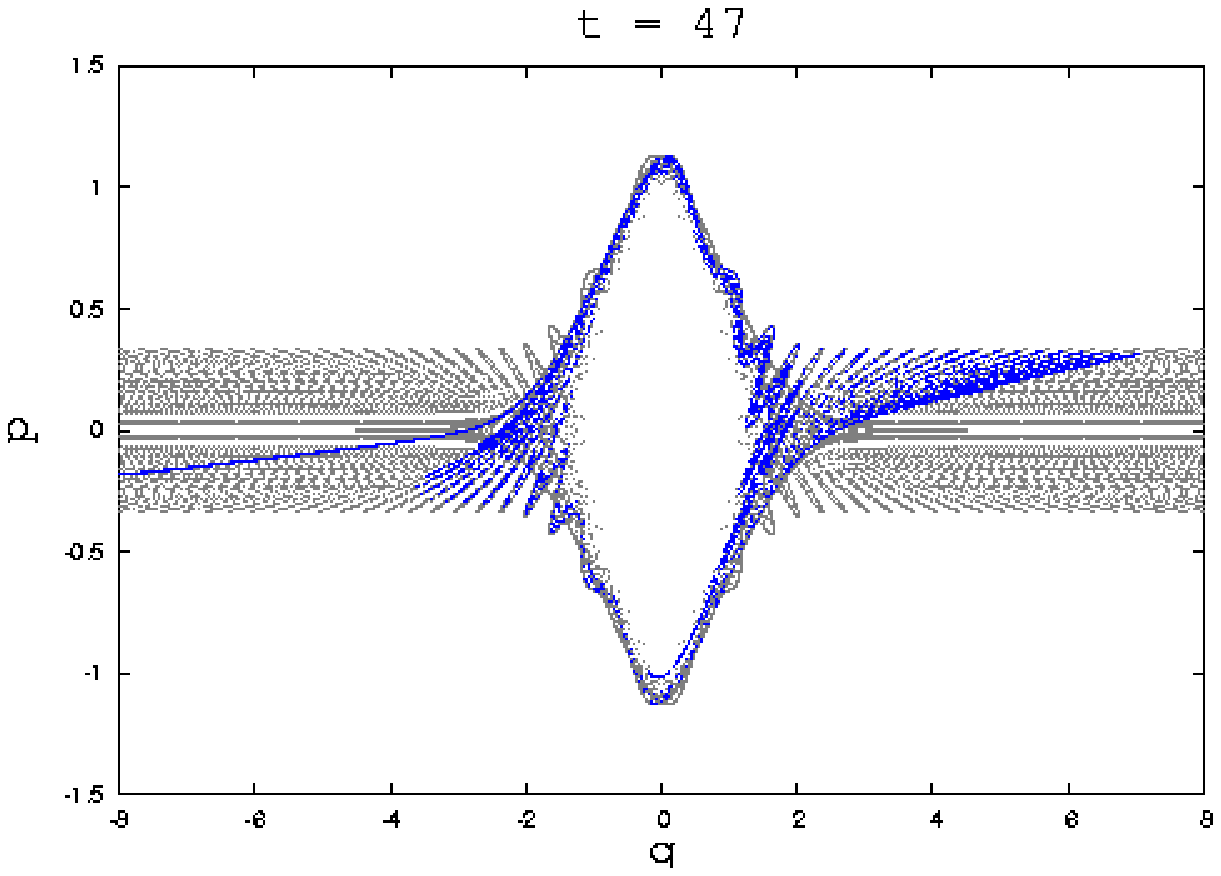}
\includegraphics[scale=.4]{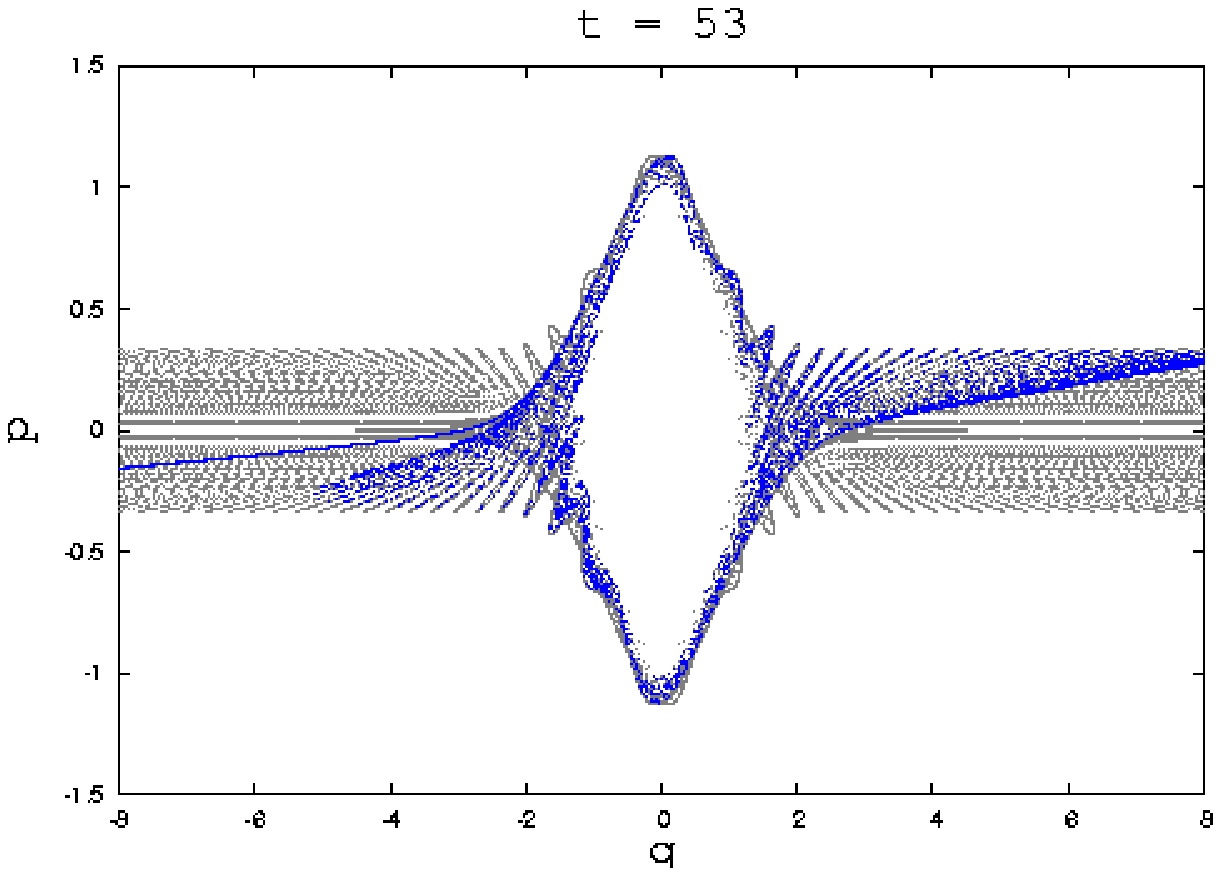}
\caption{\label{evol} Superimposed on  the horseshoe of Fig.~\ref{hor}
plotted  in grey  we  show in  red  the time  development  of a  small
particle packet in the outgoing asymptotic region and in blue the time
development  of  a small  particle  packet  placed  into the  incoming
asymptotic  region.   The  various   parts  of  the  figure  show  the
development at time  values 0, 10, 12,  20, 26, 30, 34, 42,  47 and 53
respectively.}
\end{center}
\end{figure}
In the asymptotic region the dynamics converges to free motion and the
action of the  map converges to a  pure shear motion. If we  take as a
set of initial conditions in  the asymptotic region a circular spot in
phase  space and let  it evolve  under the  asymptotic motion  then it
turns into a more and more elongated ellipse. The major axis turn more
and more horizontal under  further iteration. Figure ~\ref{evol} shows
such an initial distribution in red
and a few  of its iterated images.  In blue we see the  evolution of a
particle packet well defined in $q$  and with a fairly small spread in
$p$ starting in the incoming  asymptotic region. We see how it evolves
emitting  strongly  deformed  pulses  in both  directions  that  cover
several tendrils.

\section{Canonically transformed detectors}

\begin{figure}[!t]
\begin{center}
\includegraphics[scale=.45]{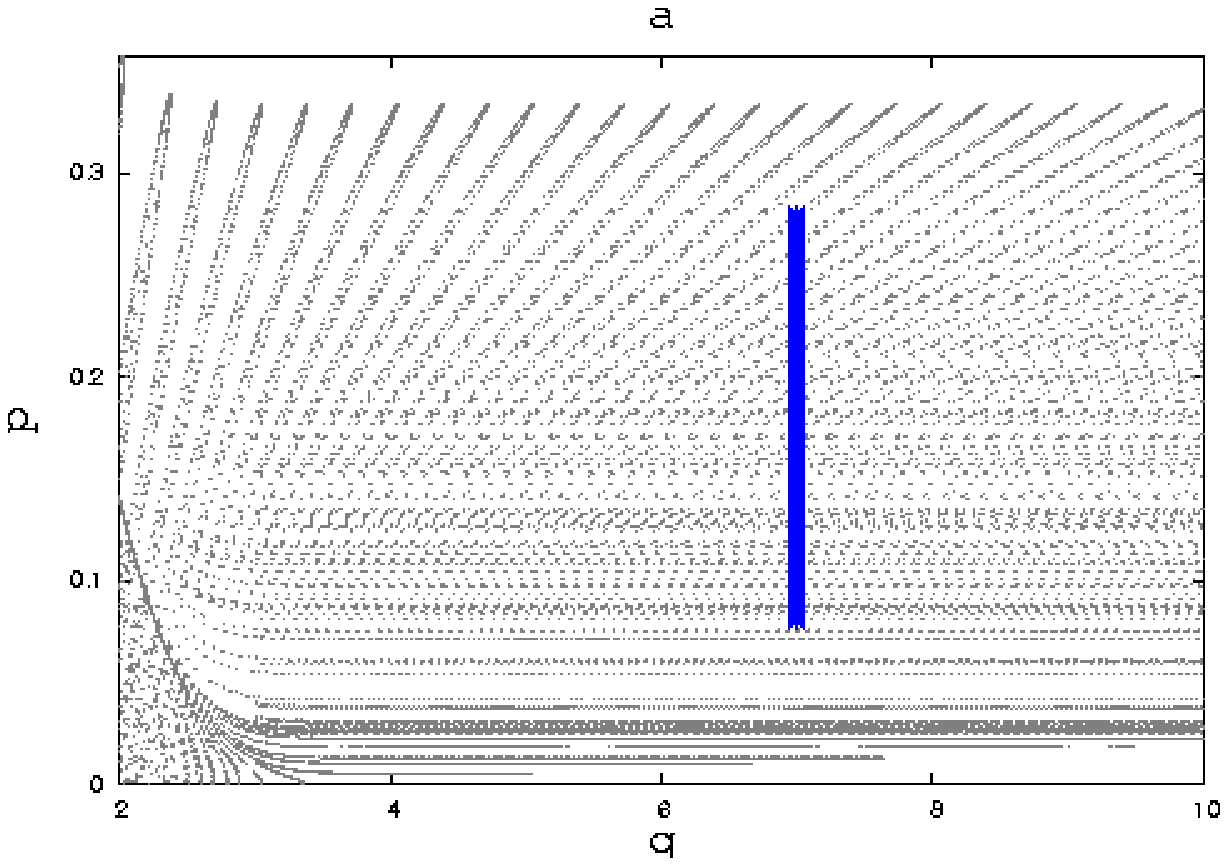}
\includegraphics[scale=.45]{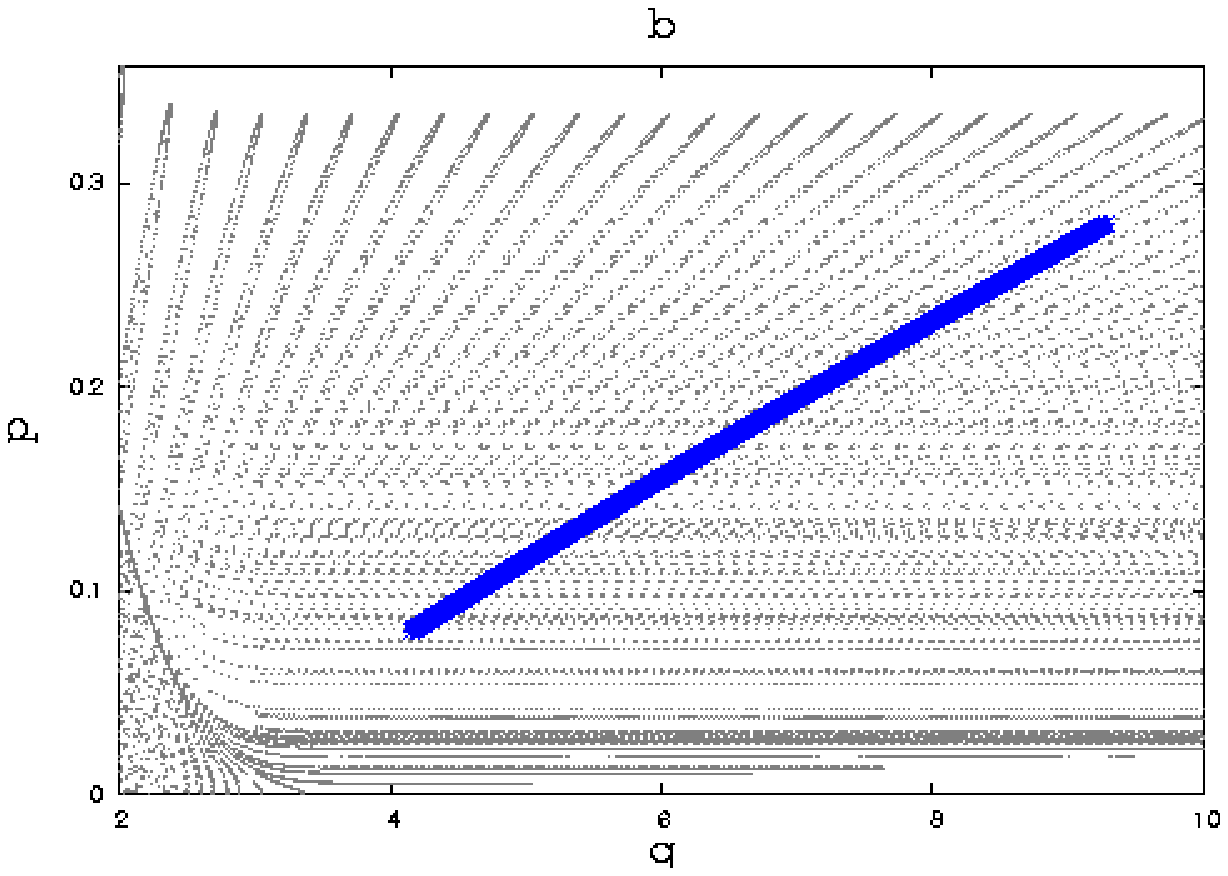}
\includegraphics[scale=.45]{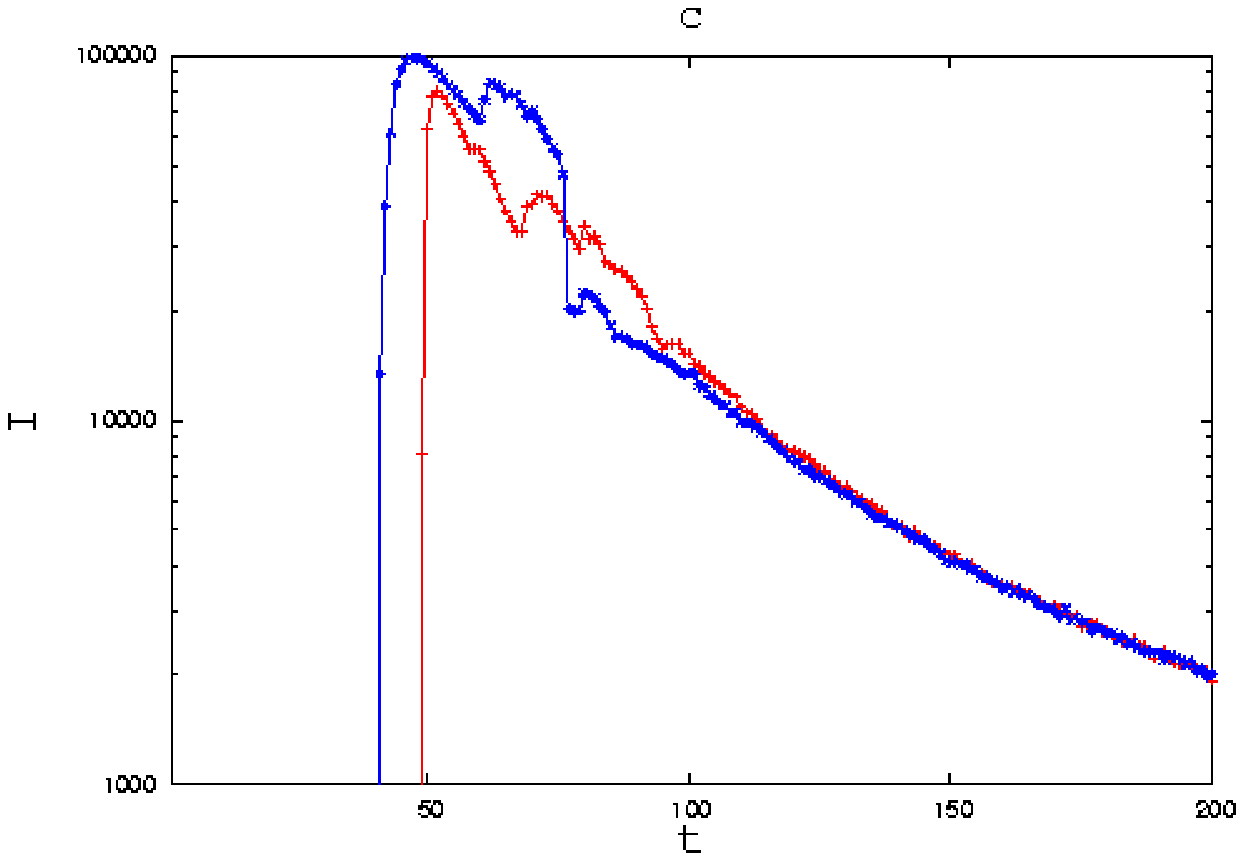}
\includegraphics[scale=.45]{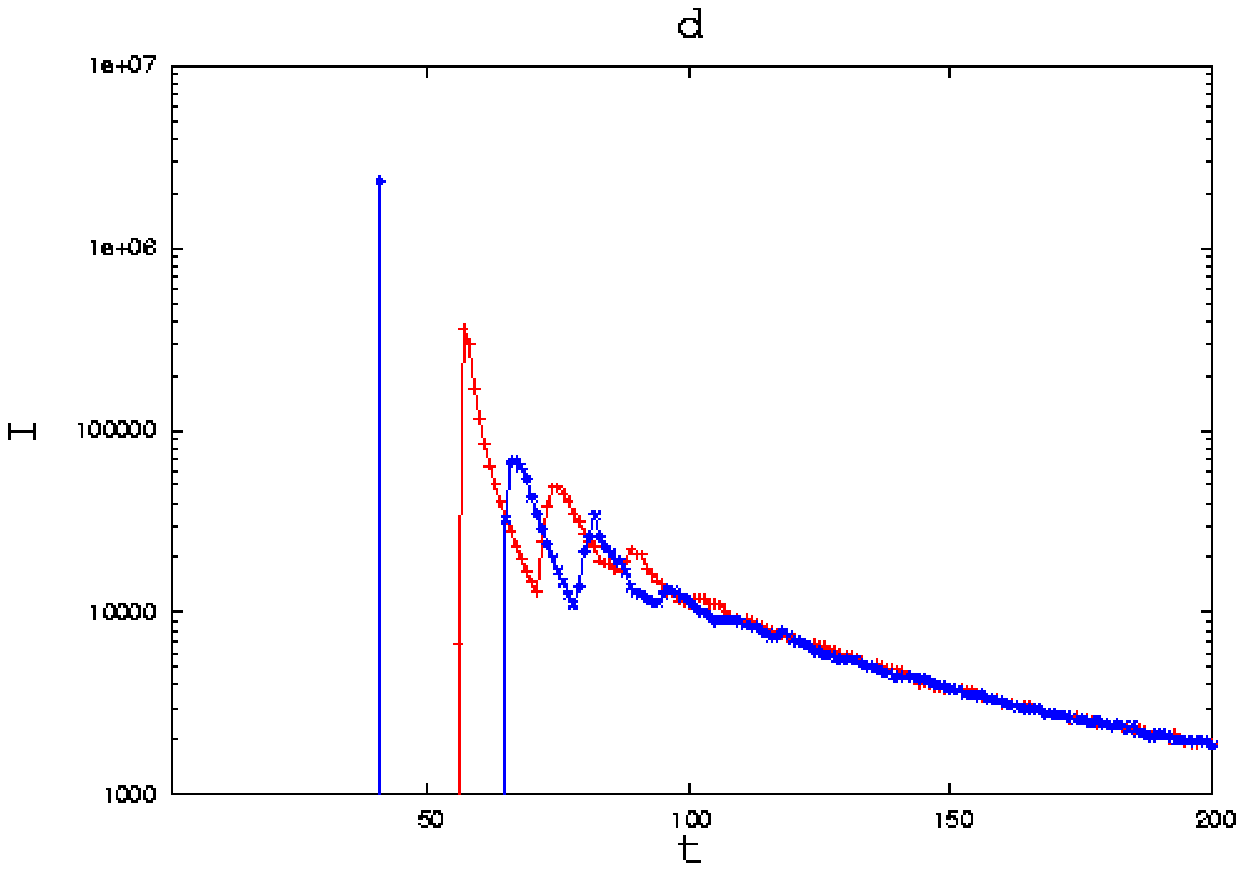}
\caption{\label{sig}The ``standard'' detector  (a) and the canonically
transformed detector (b) are  depicted together with the corresponding
number of  particle counts $I$  on a semi-log  scale as a  function of
time $t$ in (c) and (d).  The counts are integrated over one period of
the map and shown in blue for a detector on the left  and in red for a
dector on the right. The  lines are drawn  to guide the eye.  Note the
echoes in counterphase in (d)}
\end{center}
\end{figure}

Looking at Fig.~\ref{evol} we suspect,  that a detector described by a
strip cutting through  many tendrils in the outgoing  region would not
show  any  interesting features  of  the  dynamics  and certainly  not
resolve any echoes in the sense discussed in \cite{e1,darmstadt}.
This would  be the case even if  the outgoing signal would  start as a
rather clean  pulse due to the  shear seen for  the inicial conditions
marked in  red in Fig.~\ref{evol}, and  certainly for the  result of a
scattering process as marked in blue in Fig.~\ref{evol}.
Indeed  Fig.~\ref{sig}a  shows such  a  detector, and  Fig.~\ref{sig}c
shows the corresponding useless signal. The blue line gives the signal
for a detector on the left-hand side,  and the red line the one from a
detector on the right-hand side, from which we sent the incoming pulse
(c.f. Fig.~\ref{evol}).

We therefore  search for a canonical transformation  that transforms a
line of  some constant  distance in configuration  space, at  which we
would like to measure and cover an appropriate momentum interval, into
a  shape that  adjusts to  a single  tendril. In  the present  case we
choose


\begin{equation}\label{ct}
\begin{array}{l}
P=p\\
Q=q+sign(q)f(p)
\end{array}
\end{equation}
where
\begin{equation}\label{f}
f(p)=(7|p|)^{1.3}+16|p|-4.6
\end{equation}
The transformation only makes  sense in the  asymptotic region  and is
only   used  there.   The  corresponding   detector  is   depicted  in
Fig.~\ref{sig}b,  and in  Fig.~\ref{sig}d  we can  now clearly  detect
three echoes. As expected the echoes  to the right and to the left are
in counter phase.

We thus clearly  see that the CTD produced a  useful signal, where the
original detector did  not. Yet we may ask, why we  do not see signals
as neat and numerous as in ref.~\cite{e2}. There the incoming particle
pulse was launched  around the outer branch of  the stable manifold of
the fixed point. This branch does  not exist in the present case and a
launch near  the inner branch of  the unstable manifold  does not make
much sense  due to the  very low momenta.   Yet we could use  the same
procedure as for the detector and create an incoming packet that would
mimic the shape of a tendril and furthermore be sufficiently narrow to
lie within one or a  few intervals of continuity. This would obviously
produce the  desired neat image. Yet having  a canonically transformed
initial packet and detector is fortunately more then we really need to
see the echoes. But we should keep in mind, that we can operate on the
pulses or on the detectors.

\section{Discussion and final remarks}

We have seen, that CTD's, as obtained by canonically transforming more
conventional detectors  are very  helpful to produce  scattering data,
that allow to understand the inverse scattering problem.

As   an  example   we  used   a  periodically   kicked  system   in  a
one-dimensional  position  space.  Generalization  to  any  scattering
system with two important degrees of freedom where one of them is open
and  the  other  one   closed  result  trivial  \cite{i2}.  In  truely
two-dimensional systems we have to construct an appropriate Poincar\'e
section that allows the relevant horseshoe construction \cite{i1,i3}

Here  we  must recall  that  probably  the  main applications  of  the
classical  inverse   scattering  problem  arise  in   the  context  of
semi-classical  behaviour  of  wave   and  quantum  systems,  as  best
exemplified  by  an  experiment   devised  explicitly  for  this  goal
\cite{darmstadt}.  Apart  from   this  specially  designed  experiment
several  practical  situations  occur.  Examples  are:  The  collinear
scattering  of  a particle  from  a  two  particle bound  state  where
collisional  excitation of  the vibration  occurs; a  system  with any
periodic drive  by an external field (laser  assisted collisions); the
motion of an  electron in a channel {\it e.t.c.}.  In \cite{i2} it has
become evident  that we  need a detector  which measures  position and
momentum simultaneously to extract information for the inverse problem
in  the  case of  collisional  excitations  of  the closed  degree  of
freedom, but from  the considerations in the present  paper it is also
clear,  that  the  shear   in  outgoing  asymtotic  motion  will  make
canonically  transformed  detectors  necessary  even in  systems  that
display echoes  in the  near field  as soon as  we move  the detectors
sufficiently far away.

Yet the most important perspective  is the option to realize this idea
in quantum  optics, where overlaps  with coherent states  are commonly
considered as measurements  \cite{schleich}, and overlap with squeezed
states  is   an  obvious  extension;  yet  the   posibility  of  using
canonically  transformed coherent  states  in a  more general  setting
opens new perspectives, which we plan to explore in a future paper.

\section*{Acknowledgments}

The authors acknowledge  helpful discussions with C. Mejia-Monasterio,
W.   Schleich and  V.  Manko as  well  as financial  support by  DGAPA
project IN-101603 and CONACyT project 43375-F.

\label{lastpage}
\end{document}